\documentclass[unsortedaddress,onecolumn, nofootinbib,11pt]{revtex4}
\usepackage[utf8]{inputenc}
\usepackage{amsmath,empheq,color}
\usepackage{amsfonts}
\usepackage{amsthm}
\usepackage{amssymb}
\usepackage{graphicx}
\usepackage{hyperref}
\usepackage[normalem]{ulem}
\usepackage{epigraph}
\usepackage{mathrsfs}
\usepackage{bbm}
\usepackage{wrapfig}
\usepackage{lineno}
\usepackage{scalerel,stackengine}
\usepackage{epigraph}
\usepackage{cancel}
\usepackage{soul}
\usepackage{ulem}
\usepackage{epigraph}

\newtheorem{remark}{Remark}

\newtheorem{example}{Example}
\newtheorem{definition}{Definition}

\newtheorem{observation}{Observation}

\begin{document}

\title{The EPR paper: a pedagogical approach}

\author{Angel Garcia-Chung}
\email{alechung@xanum.uam.mx} 
\affiliation{Departamento de F\'isica, Universidad Aut\'onoma Metropolitana - Iztapalapa\\
San Rafael Atlixco 186, Ciudad de M\'exico 09340, Mexico}
\affiliation{Universidad Panamericana, \\ Tecoyotitla 366. Col. Ex Hacienda Guadalupe Chimalistac, C.P. 01050 Ciudad de M\'exico, Mexico}

\begin{abstract}
On the seminal paper written by Einstein, Podolsky and Rosen \cite{einstein1935can}, a critique to the completeness of quantum mechanics was posed. Part of the critique consisted in the following argument: {\it if quantum mechanics is complete, then, two physical quantities, with non-commuting operators, can have simultaneous reality.}  In this paper I aim to provide a pedagogical approach to the notions used in the EPR's argument.
\end{abstract}

\maketitle

\section{Introduction}
\epigraph{ [...] {\it the best possible knowledge of a whole does not necessarily include the best possible knowledge of all its parts; even though they may be entirely separated and therefore virtually capable of being ``best possibly known''}[...] E. Schr\"odinger \cite{schrodinger1935discussion}.}

In 1935, A. Einstein, B. Podoslky and N. Rosen published a paper \cite{einstein1935can}, with a critique to the theory of quantum mechanics. Since then, this paper became a cornerstone when understanding the principles of quantum mechanics is about. In \cite{einstein1935can}, attention was paid to the question of whether quantum mechanics can be a complete theory of physical reality or not. Einstein, Podolsky and Rosen (EPR), clearly emphasized this with the title of the paper:{\it ``Can Quantum-Mechanical Description of Physical Reality Be Considered Complete?.''}  They claimed that {\it every physical theory, must be complete}, and then, according to their analysis, they showed that quantum mechanics is not complete. To show this, they used four criteria, which I describe here as {\it completeness, elements of physical reality, locality and the uncertainty principle.} 

It is worth to mention that plenty of papers have been written discussing the concepts, the physical implications and the mathematics used in this paper but also its philosophical and the historical aspects. In this note, however, I will provide a pedagogical approach regarding these concepts based on a revision to the EPR paper. To do so, I will discuss each of the main aspects regarding the aforementioned criteria. The hope is that this note can (possibly) help undergrad physics students (and non-physicist readers) to grasp some of the insights of the EPR paper and its implications on the fundamentals of quantum physics.

 In section \ref{Completeness} we introduce the notions related with the EPR necessary criterion for a physical theory to be complete. The second criterion, {\it elements of physical reality}, is discussed in section \ref{elementsofPR}. Section \ref{LocalityandUP} shows the notions related with locality and in Section \ref{TheUP} we show its implications in the uncertainty principle of quantum mechanics. In section \ref{discussion} we provide some comments regarding the current status of the EPR analysis.

\section{Necessary criterion for a complete theory} \label{Completeness}

For EPR, the first argument to be considered is the completeness criterion. In EPR terms, {\it  ``[...] Whatever the meaning assigned to the term complete\footnote{It is referred to the concept of {\it complete physical theory}.}, the following requirement for a complete theory seems to be a necessary one: every element of the physical reality must have a counterpart in the physical theory.''} Here, the authors do not provide a definition for a theory to be complete but only a necessary condition. EPR assumed that quantum mechanics (QM) is complete, hence based on this assumption, QM should provide a physical counterpart to every element of physical reality.

 The following scheme might help to visualize the argument
$$
\begin{array}{c} \mbox{A theory is} \\ \mbox{complete if:} \end{array} \quad \left\{  \begin{array}{cc} (1) & \mbox{Every element of physical reality has a physical counterpart} \\ (2) & \mbox{other criteria}\end{array} \right.
$$
As can be seen, the criterion (1) is a necessary criterion for a theory to be complete. Once this  criterion has been stablished (and accepted), then EPR aimed to show that if QM assigns an element of physical reality to every physical counterpart, then the uncertainty principle is violated. Thus, a natural question immediately arises here: what is the definition of ``element of physical reality''? 

Due to its relevance for the EPR construction, let us expand on this concept in the next section.

\section{Elements of physical reality}  \label{elementsofPR}

In \cite{einstein1935can}, the criterion named ``element of physical reality'' plays a central role in the arguments used to consider whether quantum mechanics is complete or not. We will use this criterion  in the form of the following definition
\begin{definition} \label{DefPhysReal}
 If, without in any way disturbing a system, we can predict with certainty (i.e., with probability equal to unity) the value of a physical quantity, then there exists an {\bf element of physical reality} corresponding to this {\bf physical quantity} \cite{einstein1935can}. 
 \end{definition}

This is the exact quote of EPR's definition of {\it element of physical reality} for a given physical quantity. Let us provide some examples to illustrate this criterion, which is now a definition for us here.

\begin{example}  \label{Example1}
Consider a state of the form
\begin{equation}
| \Phi \rangle = | 1 \rangle, \label{Ex1State}
\end{equation}
\noindent where $| 1\rangle $ is some eigenfunction of the Hamiltonian operator $\widehat{H}$, i.e., 
\begin{equation}
\widehat{H} | \Phi \rangle = \widehat{H}  | 1 \rangle = E_1  | 1 \rangle .
\end{equation}
\noindent Then, according to quantum mechanics postulates, we can predict that, if the physical quantity $\widehat{H}$ is measured then its outcome, with full certainty, will be $E_1$. Hence, according to the EPR criterion \ref{DefPhysReal}, there exists an element of physical reality, which we call {\bf energy}, corresponding to the {\bf physical quantity $\widehat{H}$}. 

\end{example}

Note that we can wake such a prediction without disturbing the system, and that the mechanism by which we make such a prediction is based on the postulates of quantum mechanics rather than any measurement. In other words, if the state to be considered is an eigenstate of a given operator, then QM states(predicts) that the eigenvalue associated with the state will be the outcome of the measurement of such a physical quantity. 

Let us now consider an example where the criterion does not hold.

\begin{example}

Take a state of the form 
 \begin{equation}
| \Phi_2 \rangle = \frac{1}{2} | 0 \rangle + \frac{\sqrt{3}}{2} | 1 \rangle , \label{Ex2State}
 \end{equation} 
 \noindent where $| 0 \rangle$ and $ | 1 \rangle $ are orthonormal eigenfunctions of the Hamiltonian operator $\widehat{H}$ with eigenvalues $E_0$ and $E_1$, respectively. The state (\ref{Ex2State}) is normalized, i.e., $ | \langle \Phi_2  | \Phi_2 \rangle |^2 =  \left( \frac{1}{2}\right)^2 + \left( \frac{\sqrt{3}}{2}\right)^2 = 1$.
 
  Then, according to the postulates of QM, one fourth of the time we should get the eigenvalue $E_0$ and three quarters of the time we should get the eigenvalue $E_1$. Of course, averaging, the expected value for the Hamiltonian is $\langle \widehat{H} \rangle_{ | \Phi_2 \rangle} = \frac{E_0 + 3 E_1}{4}$. In this case, due to the state $| \Phi_2 \rangle$ is not an eigenstate of $\widehat{H}$, QM does not provide a mechanism from which we can predict the outcome of the measurement of $\widehat{H}$.  Sometimes we will get $E_0$ but others we will get $E_1$. Therefore, we cannot assign  an element of physical reality for the quantity $\widehat{H}$.

\end{example}

The first example shows that QM indeed provides with elements of physical reality to the quantity $\widehat{H}$ as long as the state is an eigenstate of $\widehat{H}$. The second example shows that QM fails to do so if the state is not an eigenstate of $\widehat{H}$. Regarding the EPR paper, that QM fails to provide an element of physical reality to the state $| \Phi_2 \rangle$ is not an issue. In \cite{einstein1935can} the attention was paid to a different route of analysis which we will continue explaining below in Section \ref{LocalityandUP}.  What is relevant for the present discussion is the ``existence of such a mechanism from which we can assign elements of physical reality to a given quantum state and its interpretation''. 

At the risk of confusing readers, let us dig a little deeper into the interpretation of this term {\emph {elements of physical reality}} in light of what we showed with the examples. To do so, let us formulate the following question: If QM cannot predict the value of the energy of the state $| \Phi_2 \rangle$, can we say then that this state has something called energy?

The standard way in which QM is taught in physics classes yields the following approach to this question: {\it if the state is an eigenstate of the Hamiltonian operator $\widehat{H}$, then the state ``has'' a defined energy. If it is not an eigenstate of $\widehat{H}$, then the state ``has not'' a  definite energy.} According to this interpretation, it seems that both states $| \Phi \rangle$ and $| \Phi_2 \rangle$ have their respectives ``energies'' but one has a ``defined energy'' while the other one ``does not have a definite energy''. However, one may ask, what is the meaning of the term ``not definite energy''?\footnote{In my opinion, with this answer the analysis is now transformed into an ambiguous semantic issue rather than a physical one.}

\begin{figure}[ht!]    
         \includegraphics[width=0.4\textwidth]{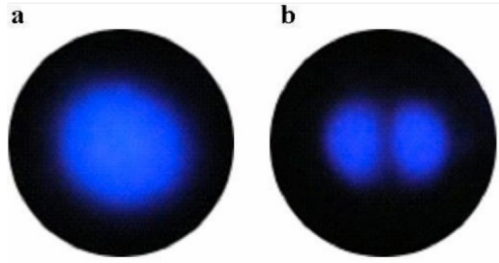}
        \caption{{\footnotesize{``Clouds of electrons surrounding carbon atoms.''}}} \label{ElectronsCloud}
\end{figure}

To clarify this point, let us move from the analysis of the energy operator $\widehat{H}$ to the position operator $\widehat{\vec{r}}$ and let us consider an image like the one in Figure \ref{ElectronsCloud} (see \cite{mikhailovskij2009imaging}).

The state of the electron on these pictures are not eigenstates of the position operator. Therefore, using the previous argument, the electrons on this figure ``do not have a definite position''. What is the meaning of this phrase? Are the electrons matter waves spread out in space and only when they are observed will they acquire the physical quantity that we name as position? or are they very well located with very well defined positions but QM fails to tell us what those positions are? Does the position of the electrons on this figure have a physical reality independently of a measurement?

According to EPR and the criterion in (\ref{DefPhysReal}) we cannot say whether there is something called ``position'' for these electrons, i.e., there is no element of physical reality associated with the position. We cannot tell, by using the wave function and with 100 \%  certainty, what the position of the electron is without disturbing in no way the quantum system. 

Is this relevant for the train of thoughts used in \cite{einstein1935can}?, the answer is no. However, it becomes problematic when interpreting some of the fundamental laws of physics as we show in the next 
\begin{remark} \label{Remark2}
For states like \ref{Ex2State} QM fails to provide an ``explanation'' about the Conservation Laws of Physics.

Consider again the state \ref{Ex2State}, then, if there is no energy associated to the state $| \Phi_2 \rangle$, is there something like the energy conservation law for this system?. 

 To illustrate this with more detail, let us say that we performed an experiment and the data collected from the measurements are given in the following table
 
 \vspace{0.5cm}
 \begin{table}[hbt!] 
 \begin{tabular}{| c | c | c | c | c | }
 \hline
 Number of measurement & 1 & 2 & 3 & 4   \\
 \hline 
 Outcome of the measurement& $E_0$ & $E_1$ & $E_1$ & $E_1$  \\
 \hline
 \end{tabular} \caption{This table contains an example of the data collected in four measurements.} \label{TableEx2}
 \end{table}
 
Recall that the initial state is $| \Phi_2 \rangle $ which does not have a definite energy value. If the state does not have a definite energy, what is happening with the energy conservation law? what ``energy'' is the one that is conserved for the state $| \Phi_2 \rangle $? $E_0$?, $E_1$?.

Note that we cannot use the statistical value $\langle \widehat{H} \rangle_{ | \Phi_2 \rangle}$. In such a case, the question that arises is: what is the meaning of the ``conservation of the statistical value of eigenvalues?''.  

\end{remark}

These questions are still open (even in the context of the so called Copenhagen interpretation) and {\it must} be addressed to better understand the fundamentals of QM. More details about the implications of this issue can be found in \cite{maudlin2020status}. Here, we will restrain ourselves of deviating our narrative in this direction and will maintain our focus on the EPR discussion. This remark is just to emphasize some of questions arising from the reality criterion given by EPR.

Now that we know the physical reality criterion used by EPR (and some of its implications) let us move to the third and fourth points on our analysis: EPR locality and the uncertainty principle.

\section{EPR Locality notion} \label{LocalityandUP}

In this section, we will show some of the details regarding the notions of locality used in EPR's paper. To do so, we will first introduce an example to illustrate part of the intuition related with the locality notion used by EPR.  
\begin{example} \label{BlocksExample}
Let us consider a classical system given by two identical blocks and a spring. In the first stage, the two blocks compress the spring as shown in Figure \ref{TwoBlocks}. In the second stage, the blocks are released and then move apart with the same speed. One goes to the left (say A) and the other block goes to the right (say B).
\begin{figure}[h!]    
         \includegraphics[width=0.4\textwidth]{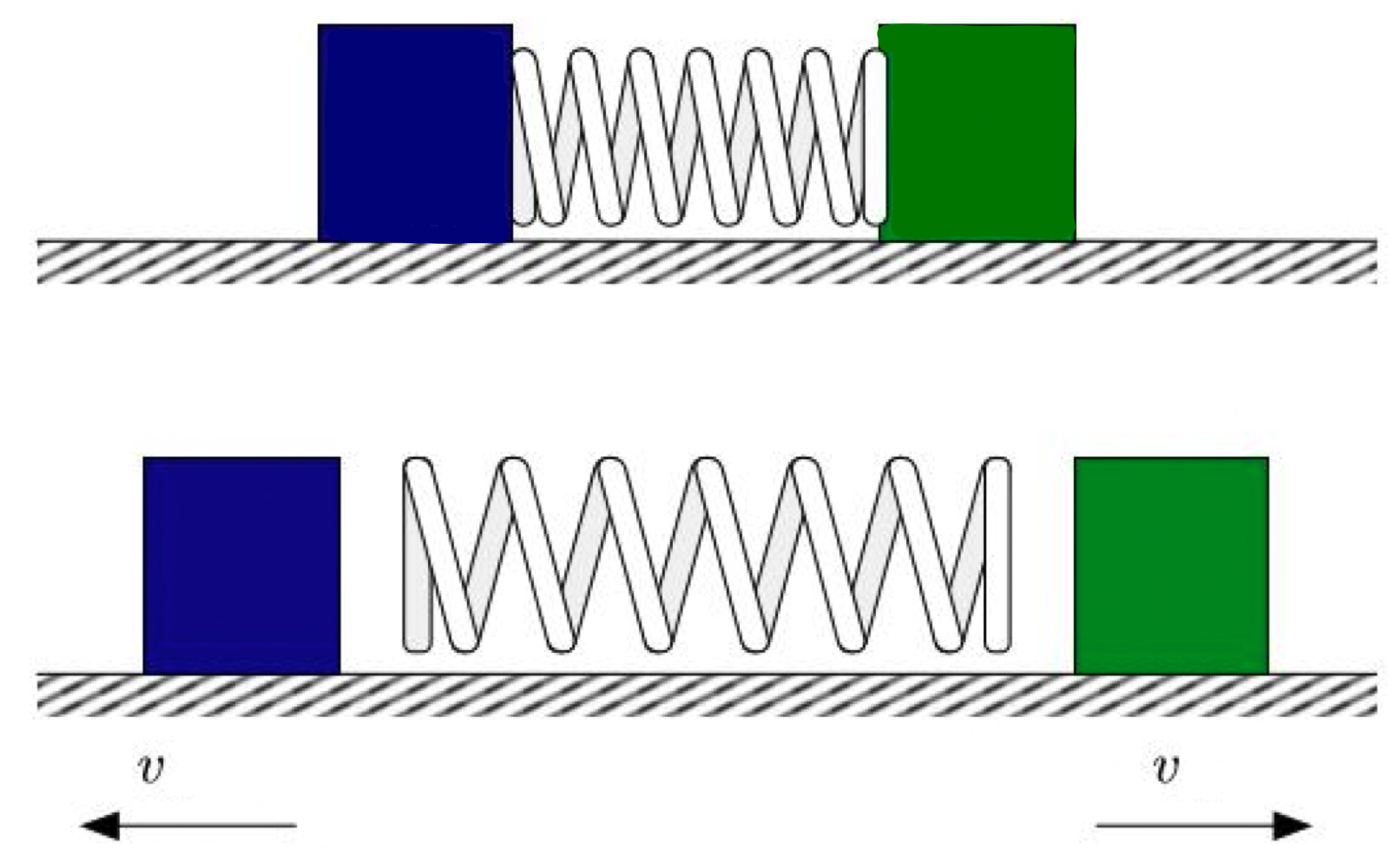}
        \caption{{\footnotesize{Two blocks that were compressing a spring are released from rest. One goes to the left (say block A) while the other goes to the right (say block B). The blocks are identical.}}} \label{TwoBlocks}
\end{figure}

Let us assume that after a while the spatial separation between the blocks is considerably large. In this way, there is no chance for the block A to gravitationally (or by any other field) interact with the block B. Now consider for instance that we lost visual contact with the block B and only the block A is in our ``vision'' (measurement) range.

 According to classical mechanics, we have a procedure to predict the position of the block B with certainty and without affecting it: just measure the position\footnote{Let assume for simplicity that the origin is the center of the spring in the first stage.} of block A, say $x_A$, and then, the position of block B will be $x_B = - x_A$. Then the definition in \ref{DefPhysReal}  implies that there exists an element of physical reality for the position of the block B, i.e., block B has a {\it definite position} even if we do not observe it.
 
  We can repeat this analysis for the momentum, the energy and the other physical quantities and we will get a definite answer for each of these quantities. All of them are elements of physical reality for the block B. Classical mechanics provides a mechanism to predict these quantities.
 
\end{example}

Now, we may ask the following questions: does the measurement of position $x_A$ cause the block B to acquire this property?, or ``was there'' the position $x_B$ for the block B before the measurement of $x_A$ or was a result of this measurement?  or in other words, does the block B have a definite position even if we do not look at it?

According to classical mechanics and the locality notion of EPR, {\it the measurement on block A could not have caused any property for block B. Thus if the position of block B is known to have this property after the measurement, it must have possessed it all along, independent of the measurement made on block A.} 

This is what we can consider as the locality notion of EPR which we present as a definition
\begin{definition} \label{DefLocality}
Whatever action on block A is a {\bf local action} and {\bf cannot affect} block B as long as they are spatially separated. 
\end{definition}

This implies that any measurement on block A do not causes block B to acquire any property, for block B was not disturbed in any way by the measurement on block A. Thus if block B is known to have a given property after the measurement, it must evidently have possessed it all along, independent of the measurement made on block A.
 
 Let us now move to the analysis of locality used in EPR but applied to QM and let us now consider the eigenstates\footnote{For completeness, I provided in the appendix the formal proof that these states are eigenstates of the position operator.} of the position operator $\widehat{x}$, which can formally be written in bra-ket notation as
\begin{equation}
| \delta_q \rangle = \int d \tilde{x}  \, | \tilde{x} \rangle \, \delta(\tilde{x} - q ). \label{KPosition}
\end{equation}

Let us now use these eigenstates to construct a state given by
\begin{equation}
| \Psi \rangle =  \int d q \; \Psi(q) | \delta_q \rangle_1 \otimes | \delta_{q + d} \rangle_2 , \label{Ex4State}
\end{equation}
\noindent where the eigenstate index  for the particle 2 is shifted in $d$ units with respect to the eigenstate of particle 1. Here $d$ is a positive real number representing, as we will see further, the spatial separation between particle 1 and particle 2. The function $\Psi(q)$ is the expansion coefficient for this state and is such that
\begin{equation}
\int dq\,  | \Psi(q) |^2 =1.
\end{equation}
\noindent This implies that the state (\ref{Ex4State}) is Dirac-delta normalized.

Let us now measure the position of particle 1 and say that the result of this measurement is that particle 1 was detected in the position $Q_1$. This implies the following collapse scheme
\begin{eqnarray}
| \Psi \rangle =  \int d q \; \Psi(q) | \delta_q \rangle_1 \otimes | \delta_{q + d} \rangle_2 \quad \rightarrow \quad | \Psi \rangle_{\rm Collapsed} &=&  \int d q \; \Psi(q)  \delta(q - Q_1)  | \delta_{q + d} \rangle_2 , \nonumber \\
&=&  \Psi(Q_1)   | \delta_{Q_1 + d} \rangle_2.
\end{eqnarray}
\noindent Consequently, the collapsed state $ | \Psi \rangle_{\rm Collapsed}$ is an eigenstate of the position operator $\widehat{x}_2$ with eigenvalue $Q_2 = Q_1 + d$. That is to say, the separation of both particles will always be the distance $d = Q_2 - Q_1$, independently of the outcome for $Q_1$. In this way, we constructed a state $| \Psi \rangle $ where the spatial separation, given by the parameter $d$, is guaranteed. 

Let us now consider that the value of $d$ is large enough to avoid any ``influence'' of particle 1 in the outcome of the position measurement of particle 2. Similar to the Example (\ref{BlocksExample}), and using the definition (\ref{DefLocality}), whatever action on particle 1 is a local action and cannot affect particle 2 as long as they are spatially separated. This is the locality notion used by EPR in \cite{einstein1935can} and note how familiar is this result for most of the physicist community: nothing we do in the moon, upon the block A, will alter the state of block B. 

Now comes the twist:
\begin{observation} \label{Observation}
Using EPR criterion (\ref{DefPhysReal}) we observe the following: We started with the state $| \Psi \rangle $ and then provided a mechanism, using QM postulates, from which we were able to predict the position of particle 2, with 100\% certainty and without affecting it. Therefore, the position $Q_2$, whatever value it takes, is an element of physical reality for the state $| \Psi \rangle$, or in other words, according to this mechanism, particle 2 has a very well defined position whose value is $Q_2$.
\end{observation}

As we will see in the next section, EPR used this result to obtain a contradiction using the Uncertainty Principle.

\section{The contradiction with the uncertainty principle} \label{TheUP}

Let us consider the eigenstates of the momentum operator given by
\begin{equation}
| p \rangle = \frac{1}{\sqrt{2 \pi \hbar}} \int dx \, | x \rangle \, e^{ \frac{i p \, x}{\hbar}} ,
\end{equation}
\noindent where $\sqrt{2 \pi \hbar}$ is a normalization factor in order to make the basis $\{ | p \rangle \}$ Dirac-delta orthonormal
\begin{equation}
\langle p' | p \rangle = \delta(p' - p).
\end{equation}
\noindent  Following a similar procedure to the one given in the Appendix, it can be checked that
\begin{equation}
\widehat{p} | p \rangle = p | p \rangle.
\end{equation}
Let us consider then the same state $| \Psi \rangle$ in (\ref{Ex4State}) but now written in this basis as follows
\begin{equation}
| \Psi \rangle =  \int d p \; \widetilde{\Psi}(p) | - p \rangle_1 \otimes | p \rangle_2 . \label{Ex5State}
\end{equation}

 If we equate the states (\ref{Ex4State}) and (\ref{Ex5State}), we will obtain that the coefficient $\widetilde{\Psi}(p)$ is related with the coefficient $\Psi(q)$ via the following relation
\begin{equation}
\widetilde{\Psi}(p) = \Psi(x_2 -d) e^{- i\frac{d p}{\hbar}},
\end{equation}
\noindent hence, this coefficient is normalized
\begin{equation}
\int d p \, | \widetilde{\Psi}(p) |^2 = \int d q \, | \Psi(q) |^2 = 1.
\end{equation}

Recall that according to Observation (\ref{Observation}) the state (\ref{Ex4State}) is such that particle 2 has a definite position $Q_2$. Therefore, due to both, state (\ref{Ex4State}) and state (\ref{Ex5State}) are the same, then the state (\ref{Ex5State}) also has a definite position $Q_2$ for the particle 2.

Let us now measure the momentum of particle 1 and say its value is $P$. This yields the following collapse scheme
\begin{eqnarray}
| \Psi \rangle =  \int d p \; \widetilde{\Psi}(p) | - p \rangle_1 \otimes | p \rangle_2 \qquad \rightarrow \qquad | \Psi \rangle_{\rm Collapsed} &=&  \int d p \; \widetilde{\Psi}(p) \, \delta(p + P)\, | p \rangle_2, \nonumber \\
&=&  \widetilde{\Psi}(-P) \,  | -P \rangle_2,
\end{eqnarray} 
\noindent which means that the collapsed state $| \Psi \rangle_{\rm Collapsed}$ is now an eigenstate of the momentum operator of the particle 2 with eigenvalue $-P$. 

Consequently, we have provided another mechanism from which we where able to predict the momentum of particle 2, with 100\% certainty and without affecting particle 2. Therefore, according to (\ref{DefLocality}) the particle 2 has a definite momentum $-P$, but, according to (\ref{Observation}) it also has a definite position $Q_2$. This clearly contradicts the Uncertainty Principle which states that: due to position and momentum operators do not commute, then a particle must not simultaneously have well defined position and momentum.

Hence, if we assume that quantum mechanics is complete, together with the criteria used before, then it leads to a contradiction.

\section{Final comments} \label{discussion}

There is no doubt that the EPR argument was, at the time, demolishing for the foundations of QM. For example, E. Schr\"odinger was highly impressed by this paper and came with his own interpretation of it, coining, at the same time, the term {\it entanglement} \cite{schrodinger1935discussion}. N. Bohr was also very impressed by it and wrote a reply \cite{bohr1935can}, apparently, giving rise to what is sometimes known as the {\emph{Einstein-Bohr debate}} \cite{whitaker2004epr}.

  Years later, J. S. Bell analyzed the EPR argument but using a different perspective \cite{bell1964einstein}. Bell's argument is based on a theorem with severe implications in our understanding not only of the EPR argument but also on the foundations of physics, see for example, \cite{bell2004speakable}. According to Bell, if some inequalities, now known as Bell's inequalities, are satisfied, then the EPR's locality notion is valid. However, every experiment carried out to prove Bell's inequalities has shown that EPR's locality notion does not hold.
  
   The problem is then that, not being valid the EPR argument does not implies the validity of any of the alternative interpretations of reality and locality for QM (see for example \cite{ghirardi1986unified, bassi2016models, bassi2013models, norsen2006epr}). This motivates the current efforts to explain the measurement problem \cite{maudlin1995three, maudlin2011quantum}. This, however, might be the reason for our next pedagogical journey. 


\section{Acknowledgments}
I acknowledge Academia de Matem\'aticas and Colegio de F\'isica, UP, for the support and enthusiasm and also, Daniel Flores, Daniel Sudarsky and Elias Okon for their helpful comments and discussions.

\section*{Appendix}

Let us check that $| \delta_q \rangle$ is actually an eigenstate of $\widehat{x}$. First, recall that
\begin{equation}
\langle x | \delta_q \rangle = \int d \tilde{x}  \,\langle x | \tilde{x} \rangle \, \delta(\tilde{x} - q ) = \int d \tilde{x} \,  \delta(x- \tilde{x}) \, \delta(\tilde{x} - q )= \delta(x - q ), \label{WFPosition}
\end{equation}
\noindent i.e., the Dirac delta $\delta(\tilde{x} - q )$ in (\ref{KPosition}) is nothing but the wave function associated with the state $| \delta_q \rangle$. This is the reason for the notation $\delta_q$ in $| \delta_q \rangle$. Moreover, these states comprise a set of Dirac-delta orthonormalized states, i.e.,
\begin{equation}
\langle \delta_{q'} | \delta_q \rangle = \delta(q' - q).
\end{equation}

The position operator $\widehat{x}$, when acting on $| \delta_q \rangle$ gives
\begin{equation}
\widehat{x} | \delta_q \rangle = \int d \tilde{x} \, \delta(\tilde{x} - q ) \, \left( \widehat{x} | \tilde{x} \rangle \right) = \int d \tilde{x} \, \delta(\tilde{x} - q ) \,  \tilde{x} | \tilde{x} \rangle ,
\end{equation}
\noindent and multiplying this expression on both sides by the bra $\langle x|$ vector, gives
\begin{equation}
\langle x | \widehat{x} | \delta_q \rangle = \int d \tilde{x} \, \delta(x-  \tilde{x}) \, \delta(\tilde{x} - q ) \,  \tilde{x}  = q \delta(x - q ) = q \langle x | \delta_q \rangle, \label{XPos}
\end{equation}
\noindent where in the last line we inserted the relation (\ref{WFPosition}). Factorizing $\langle x |$ on both sides of (\ref{XPos}) gives
\begin{equation}
\langle x | \left(  \widehat{x} | \delta_q \rangle  -  q |  \delta_q \rangle \right) = 0,
\end{equation}
\noindent which must hold for any $\langle x |$. This leads to the following relation
\begin{equation}
\widehat{x} | \delta_q \rangle  =  q |  \delta_q \rangle ,
\end{equation}
\noindent which clearly states that (\ref{KPosition}) is an eigenstate of $\widehat{x}$ with eigenvalue $q$.

\end{document}